\documentclass[aps,prl,twocolumn]{revtex4}

\usepackage{color}
\usepackage{amsmath}
\usepackage{amssymb}
\usepackage[pdftex]{graphicx}

\begin{document} 
\title{Cosmological classicalization:\\Maintaining unitarity under relevant deformations of the Einstein-Hilbert action}
\author{
Felix Berkhahn$^{ab}$, Dennis D.~Dietrich$^{c}$, and Stefan Hofmann$^{ab}$}
\affiliation{
\hskip -1.4mm$^a$Arnold Sommerfeld Center for Theoretical Physics, Ludwig-Maximilians-Universit\"at, Theresienstra{\ss}e 37, 80333 Munich, Germany\\
\hskip -1.4mm$^b$Excellence Cluster Universe, Boltzmannstra{\ss}e 2, 85748 Garching, Germany\\
\hskip -1.4mm$^c$CP$^\mathit{\hskip 0.2mm 3}$-Origins, Centre for Particle Physics Phenomenology, University of Southern Denmark, Campusvej 55, 5230 Odense M, Denmark\\~\\
}

\begin{abstract}
Generic relevant deformations of Einstein's gravity theory 
contain additional degrees of freedom that have a 
multi-facetted stabilization dynamics on curved spacetimes. 
We show that these relevant degrees of freedom 
are self-protected against unitarity violations 
by the formation of classical field lumps that eventually
merge to a new background geometry. 
The transition is heralded by the massive decay of the original
vacuum
and evolves through a strong coupling regime. 
This process fits in the recently proposed classicalization mechanism 
and extends it further to free field dynamics on curved backgrounds.
\end{abstract}

\maketitle

\textit{Introduction.---}At the core of Einstein's gravity theory is a
democratic principle guaranteeing any source
the same coupling to spacetime, independent
of its nature. Exploiting this principle has been
the successful strategy sui generis to infer the vacuum's
energy density by observing the Universe's expansion
history.

Ever since Cosmology has advanced to a high precision
science, the window of opportunity to probe gravity's
rigidity on cosmological scales is wide open.
Deforming Einstein's theory of gravity introduces
additional relevant degrees of freedom, which have a profound impact: they violate the gravitational
coupling's universality. Such proposals might
be crucial to cure the spacetime impact of technically
unnatural sources, such as the vacuum energy density,
which is in staggering conflict with the observed
expansion history.

At the linear level, a seminal mechanism for this
is at work in the Fierz--Pauli theory for massive spin-2
excitations on a Minkowski background, where the
relevant deformation corresponds to a mass term that
is unique by consistency requirements.
On generic backgrounds, the principle of equivalence
demands
additional
geometric deformations
that allow for a richer phenomenology at every level
of the effective field theory description.

In this Letter we consider generic relevant deformations of
the Einstein--Hilbert action on arbitrary backgrounds
at the linear level. Explicit results are shown for
Friedmann--Robertson--Walker (FRW) spacetimes. We analyze the classical and quantum stability
of the `free theory,' which is a necessary prerequisite
before completing the corresponding deformations at the
nonlinear level.

We show that the stabilization dynamics for the additional
relevant degrees of freedom on a curved background,
even when they are `free,' is as multi-facetted and rich
as self-protection mechanisms in certain non-renormalizable
interacting systems on a Minkowsk geometry.

In particular, we show that the recent classicalization
proposal \cite{Dvali:2011nj,Dvali:2010ns,Dvali:2010jz,Dvali:2010bf} is at work: the dynamics of the additional
degrees of freedom protects them against unitarity violations
via the formation of classical objects that eventually become
the new background spacetime.
The transition to the new geometric ground state is heralded
by the massive decay of the original vacuum and evolves
through a strong coupling regime. 

~\\

\textit{Framework.}---The effective Lagrangian 
describing the dynamics of a single dimensionless scalar field $\Phi$ 
coupled to the metric field $\mathrm{g}$, organized as a derivative expansion,
is given by 
\begin{equation} \label{EFT}
-\mathcal{L}_{\rm eff}
= 
\sqrt{-\mathrm{g}}\sum_{n=0}^{\infty}\sum_{j=0}^{n}
M^{4-2n}C^{(2n)}_{2j}(\Phi)R^{n-j}\left(\nabla\Phi\right)^{2j} 
.
\end{equation}
For simplicity, all terms have been written down schematically
to sketch their scaling with the fiducial mass $M$.
$2n$ is the total number of derivatives at this level.    
The coefficient $C^{(0)}_0$ is the non-derivative
part of the $\Phi$ self-interactions. The characteristic
scale $v$ of this term might be well below $M$, and
$C^{(0)}_0 \propto (v/M)^4$. At $n=1=j$
the kinetic terms enter.  
$R^{n-j}$ stands for all possible combinations of a total of $n-j$ Ricci scalars, tensors, and Riemann tensors. 
Note that the fiducial mass scale $M$
could be much smaller than the reduced Planck mass $M_{\rm P}$, 
in which case $C^{(2)}_{0,2} \propto (M_{\rm P}/M)^2$. 
Further and $\Phi$ independent gravitational 
sources could be added.

For the purpose of studying the stability of the effective theory
(\ref{EFT}), we expand about classical background configurations 
and geometries, $\Phi=\Phi_0 + \phi/M$ and
$\mathrm{g}=g + h/M$. Of particular interest is the case 
where the background configuration is decoupled from
the background geometry, i.e., when $\Phi_0 \equiv 0$.
Expanding the effective theory (\ref{EFT}) up to second
order in the fluctuations $\phi$, the kinetic sector for $\phi$
becomes 
\begin{equation} \label{kin}
-2 \mathcal{L}_{\rm kin} 
=
\sqrt{-g}\; [g^{\mu\nu}+\mathcal{F}^{\mu\nu}(R/M^2)] \; \nabla_\mu \phi \nabla_\nu \phi
\; .  
\end{equation}
Generically, the matrix $\cal F$ has no definite signature and thus, 
the perturbative consistency of (\ref{kin}) is rather sensitive
to the geometrical background. 
It will prove to be useful to recast it
in terms of a canonical kinetic term and a coupling to a $\phi$ dependent source $\mathcal{J}$,
\begin{eqnarray} \label{kin2}
2 \mathcal{L}_{\rm kin}/\sqrt{-g}
&=&
\phi \; \Box \; \phi + \mathcal{J} (R/M^2, \phi) \phi
\; ,
\end{eqnarray}  
with $\mathcal{J}\equiv \nabla_{\mu} \left( \mathcal{F}^{\mu \nu} \nabla_{\nu} \phi \right)$, where potential boundary contributions have been suppressed for the moment. 

Starting from (\ref{kin}) we can analyze the stability of the system by looking after imaginary contributions to the one-loop effective Lagrangian (the vacuum persistence amplitude),
\begin{equation} \label{eff_lag}
2\mathcal{L}^{(1)}
=
\ln\mathrm{Det}\{\nabla_\mu[(g^{\mu\nu}+\mathcal{F}^{\mu\nu})\nabla_\nu\Box^{-1}]\} ,
\end{equation}
which is normalised to the free part $\Box$.
In the short-distance limit we get contributions of the form
\begin{equation}
2\mathcal{L}^{(1)}\supset\ln\mathrm{Det}[(g^{\mu\nu}+\mathcal{F}^{\mu\nu})\partial_\mu\partial_\nu\Box^{-1}] .
\end{equation}
The fact that the matrix $g^{\mu\nu}+\mathcal{F}^{\mu\nu}$ can, in general, have a signature different from the metric $g^{\mu\nu}$ can lead to negative arguments of the $\ln$ and thereby, to imaginary contributions, which signal the decay of the vacuum. 

However, here a signature change is heralded by a strong coupling regime,
and, as a consequence, the degrees of freedom that trigger the vacuum decay are 
self-protected against unitarity violation. 

There are two known protection 
mechanism between which a theory can choose to establish consistency. Either it allows
for weakly coupled heavy degrees of freedom with masses above the original 
strong coupling scale, or it generates classical field configurations 
via energy-momentum self-sourcing, corresponding to a feedback
through nonlinear terms. The latter case is the recently
proposed classicalization mechanism, see below. 

~\\

\textit{Cosmology.}---For concreteness, let us consider Fierz--Pauli theory on FRW spacetimes. 
There, five instead of two degrees of freedom propagate. 
One of these supplementary degrees of freedom, which, in the spirit of the Goldstone boson equivalence theorem dominates the dynamics at high energies 
corresponds to the field $\phi$. As a consequence, Fierz--Pauli theory fits in the framework (\ref{EFT}). To be more precise, the metric fluctuation $h_{\mu \nu}$ is written as
\begin{equation}
h_{\mu \nu} = \tilde{h}_{\mu \nu} + \nabla_{(\mu} A^{T}_{\nu)} + \nabla_{\mu} \nabla_{\nu} \phi
\end{equation}
where $\tilde{h}_{\mu \nu}$ carries two degrees of freedom, like the massless graviton would propagate. This clarifies how $\phi$ feeds into the spacetime fluctuations. 

In Fierz--Pauli theory over an FRW spacetime we find for the source \cite{Berkhahn:2010hc}
\begin{equation} \label{curr_FRW}
m^2\mathcal{J} =(\dot H + H^2) \ddot \phi + (\ddot H + 5 H \dot H + H^3) \dot \phi + (\dot H + 3 H^2) \frac{\vec{\nabla}^2}{a^2} \phi \; ,
\end{equation}
where $H=H(t)$ denotes the Hubble parameter, $a=a(t)$ the scale factor and $m$ the deformation parameter of Fierz-Pauli theory, which would be interpreted as the graviton mass on a Minkowski background. Moreover, 
$\mathcal{F}^{\mu \nu}(R/M^2) = - R^{\mu \nu}/3m^2$.
Then stability of the vacuum state requires
\begin{equation} 
\label{stability_bound}
m^2 > H^2 + \dot H/3 \; .
\end{equation}
We had already discovered this bound in \cite{Berkhahn:2010hc}, applying a classical stability analysis to the system (\ref{kin}) for the case of Fierz--Pauli gravity. 
This classical stability bound (\ref{stability_bound}) arises
when the spatial components of $g^{\mu\nu}+\mathcal{F}^{\mu\nu}$ change sign.
In fact, the coefficient in front of $\vec\nabla^2\phi/a^2$ in (\ref{curr_FRW}),
which coincides with $\mathcal{F}^{ii}$, changes its sign relative 
to $g^{ii}$.

\begin{figure}[!htbp]
  \centering
\includegraphics[width=8.6cm]{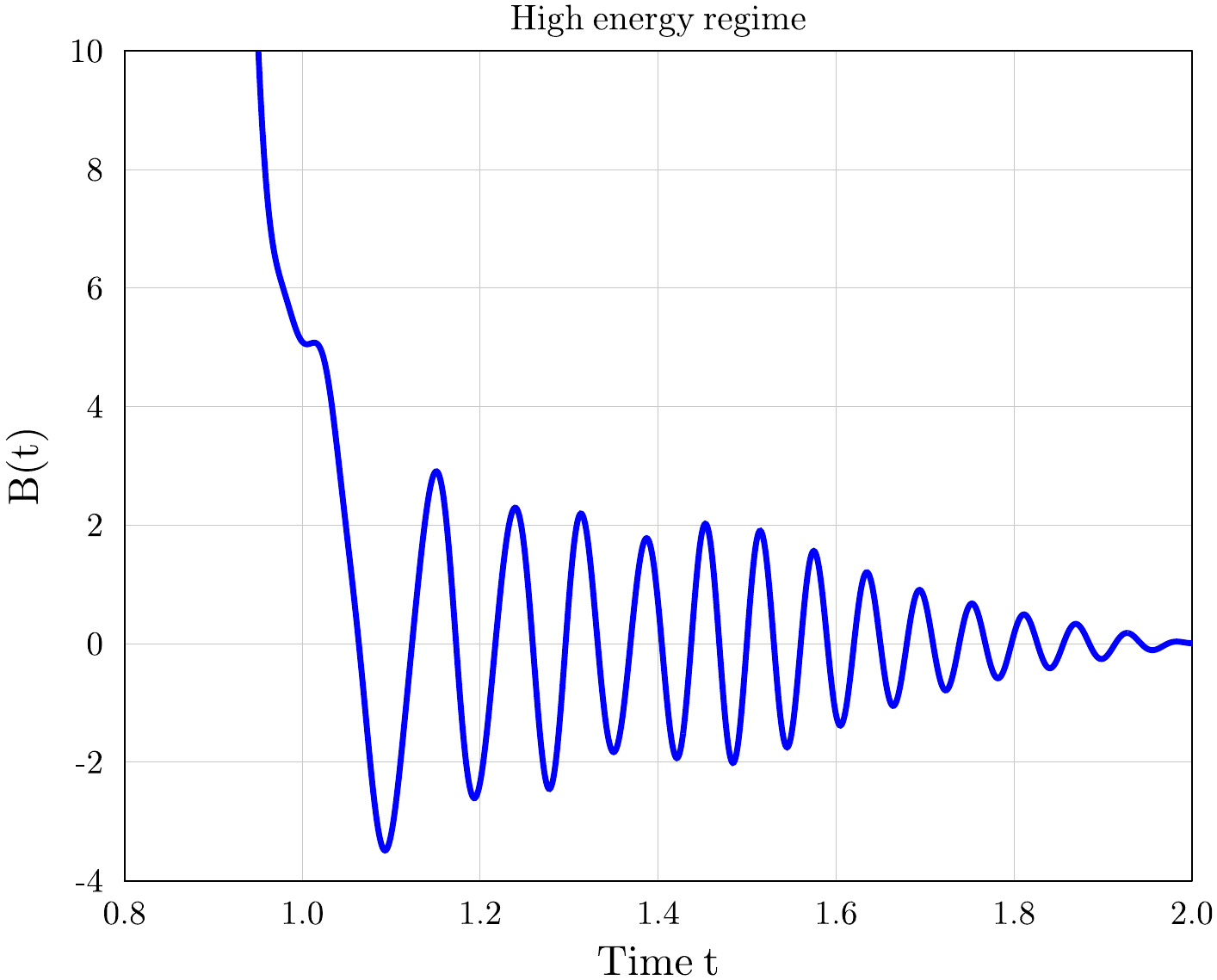}
\caption{Numerical analysis in the high energy regime for a radiation dominated universe \cite{Berkhahn:2011hb}. $m$ is chosen such that the bound (\ref{stability_bound}) is violated for $t<1$.  
$B$ corresponds to $\phi$.}
\label{fig_high_energy}
\end{figure}

A violation of the stability bound manifests itself in an explosion of the otherwise oscillating fluctuation solution. This is shown in Fig.~\ref{fig_high_energy}, where $B$ corresponds to $\phi$. 
The loss of classical stability typically signals that the system evolves into a new classical background solution. Equipped with this knowledge, we are able to reinterpret the above decay of the vacuum: Many  
free modes 
pop out of the vacuum and their superposition yields a 
classical object with large occupation number. In this case,
the largest possible classical object is formed ---
a new spacetime background. 

At this level in the effective field theory description, the 
newly formed classical field configuration evolves according
to the differential operator in the numerator of (\ref{eff_lag}).
It is then decomposed into a superposition of 'free' solutions 
(those of $\Box$), which present potential decay modes.
Especially this last aspect will become important below.

Even though the vacuum persistence amplitude cares about 
quantum mechanical consistency, its non-normalizablitlity,
in our case, can be traced back solely to the background decay,
and bears no impact on the intrinsic consistency of 
the quantum theory. 
 
For the temporal components in the one-loop effective Lagrangian the sign change is postponed to
\begin{equation} \label{unitarity_bound}
m^2 > \dot H + H^2\; .
\end{equation}
This is linked to the coefficient of $\ddot{\phi}$ in the source (\ref{curr_FRW}), which coincides with $\mathcal{F}^{00}$, and which has to be compared to $g^{00}$.

In \cite{Berkhahn:2010hc,Berkhahn:2011hb}, we  
found that the relation (\ref{unitarity_bound}) must be satisfied to ensure the absence of negative norm states. For non-phantom matter ($\dot H < 0$) the classical stability bound (\ref{stability_bound}) is stronger than the unitarity bound (\ref{unitarity_bound}). Therefore, we cannot trust this derivation of the unitarity bound at all, since, whenever it looks as if the theory would contain negative norm states, it is no longer in the perturbative regime. 
There is thus the hope that a full non-linear theory might not contain any unitarity violating negative norm states. We call this self-defense mechanism of the linear theory 'self-protection' \cite{Berkhahn:2010hc}.

~\\

\textit{Classicalization.}---The above is a striking example for the concept of classicalization \cite{Dvali:2011nj,Dvali:2010ns,Dvali:2010jz,Dvali:2010bf}: Classicalization is a unitarization mechanism based on energy-momentum self-sourcing at variance with unitarization by weakly interacting short-distance physics (the Wilsonian mechanism). In a nutshell, at high energies, the formation of a classical object (the `classicalon') inhibits interactions at short distances, which leads to the unitarization of the process. 
The classical object formed in the course of the unitarization process will finally decay into a large number of final states, which is the prime signal for this mechanism.
As explained just above Eq.~(\ref{unitarity_bound}), the analysis (\ref{eff_lag}) has also exactly this interpretation.
 
For the example of Fierz-Pauli theory it can be ascertained that the number of free modes is indeed large, as the bound (\ref{stability_bound}) is independent of how short distances we regard. As a consequence, there will be contributions to the imaginary part of the determinant from arbitrarily small distances.

This brings us back to the aforementioned concept of energy/momentum self-sourcing \cite{Dvali:2011nj,Dvali:2010ns,Dvali:2010jz,Dvali:2010bf}. Self-sourcing occurs in interacting field theories in which the interaction terms contain sufficiently many derivatives. In such a setting solutions with small amplitudes but sufficiently high four-momentum lead to a strong enhancement of exactly these interaction terms and to unitarization by classicalization. In Eq.~(\ref{kin2}) this is realized in the source $\mathcal{J}$: First, there occurs a function of the curvature $R$, which power by power contains two derivatives of the background geometry, and additionally, there are the derivatives of $\phi$. Accordingly, for Fierz--Pauli gravity over an FRW spacetime, the source (\ref{curr_FRW}) contains the curvature $H$ and its temporal derivatives as well as derivatives of $\phi$. 
For rather generic choices of spacetime sources, the curvature part behaves as 
$\mathcal{J} \supset R_{\mu \nu} \propto 1/t^\alpha$ . The increase
of the self-source's strength with increasing localization 
is a feature characteristic for classicalization \cite{Dvali:2010jz}. 
In our case, due to the background isometries, the localization scale is a time-scale.

Self-sourcing does not stop there. In a next step, 
the fluctuations $\phi$ and particularly their derivatives would become 
sufficiently sizable to trigger a change of the background spacetime geometry. 

~\\

\textit{Dictionary.}---Restoring explicit insertions of $\hbar$ we are able to distinguish between mass 
and length scales. After canonically normalizing the field $\phi$, 
the parameter in front of $R^{\mu \nu} \partial_{\mu} \phi \partial_{\nu} \phi $ has units of length squared. In the same way, the parameter in front of the Fierz--Pauli combination has units of inverse length squared. 
The  corresponding term in (\ref{EFT}) should thus be written as
\begin{equation}
M^{-2} R^{\mu \nu} \partial_{\mu} \phi \partial_{\nu} \phi \rightarrow L^2 R^{\mu \nu} \partial_{\mu} \phi \partial_{\nu} \phi .
\end{equation}
In Fierz--Pauli theory, $L$ would denote the screening length of the gravitational field. 
The bound (\ref{unitarity_bound}) defines the time $t_U$ where the theory would violate unitarity, whereas (\ref{stability_bound}) defines the time $t_{\star}$ where the classical theory becomes strongly coupled. For a cosmology that is dominated by matter with the equation of state $\partial p/\partial\rho=w$, their values are given by,
\begin{eqnarray}
t_U &=& \frac{\theta({\hbar})}{3} \frac{L}{1+w} \sqrt{2(-1 - 3w)},  \label{tU}\\
t_{\star} &=& \frac{1}{3} \frac{L}{1+w} \sqrt{2(1-w)}\; . \label{tStar}
\end{eqnarray}
We have explicitly restored the $\hbar$ dependence in the expression (\ref{tU}) with the convention $\theta(0)=0$. The characteristic time scale for unitarity violation, $t_U$,
is a direct consequence of a quantum commutation relation, which explains 
the appearance of the Heaviside function $\theta$. 
The characteristic time scale (\ref{tStar}) for violating classical stability does not
contain $\hbar$, as it is solely set by classical physics. 
Given that $H^{-1}\propto t$ is the Hubble length, it is natural to reinterpret the above times as length scales, both of which are directly proportional to the Fierz--Pauli length $L$.

The existence of two time scales, one characteristic for quantum instability, 
the other for classical instability, and their hierarchy $t_\star>t_U$ 
(provided Im$(t_U)=0$), is again in analogy to the classicalization 
concept \cite{Dvali:2011nj,Dvali:2010ns,Dvali:2010jz,Dvali:2010bf}.

Note that for many reasonable values of $w$ the time scale $t_U$ is imaginary. 
In this case, the would-be unitarity bound is absent. However, for mixtures of a cosmological constant with other FRW sources, 
$t_U$ will give some finite positive number.

One might ask whether it is possible to cure the theory by a Wilsonian treatment, 
that is, integrating in new heavy degrees of freedom, instead of creating classical objects. To do so, we replace 
\begin{equation}
\mathcal{F}^{\mu\nu}(R/M^2) \; \nabla_\mu \phi \nabla_\nu \phi \rightarrow \mathcal{F}^{\mu\nu}(R/M^2) \; \frac{\Lambda^2}{\Box + \Lambda^2} \nabla_\mu \phi \nabla_\nu \phi \; .
\end{equation}
At high energies, $\Box \gg \Lambda^2$, this term will loose its
kinetic nature. Accordingly, we would be left with a standard kinetic term for $\phi$, 
and there would neither occur a stability nor a unitarity issue. 
However, in the opposite regime, $\Box \ll \Lambda^2$, 
this modification is negligible, 
and there is no window for a Wilsonian cure of the theory. 
Instead, classicalization must occur.

In particular, in the case of Fierz--Pauli theory, the Lagrangian (\ref{kin}) only describes the theory in the high energy regime, $\Box \gg m^2, H^2$. For phenomenological reasons, we typically take $m^2 \approx H^2$, so that we effectively have $\Box \gg m^2$. Moreover, any new Wilsonian heavy degrees of freedom must have a mass $\Lambda$ much above $m$, $\Lambda^2 \gg m^2$. Otherwise, the effective theory (\ref{kin}) would have been incomplete. Hence, intermediate energies $m^2 \ll \Box \ll \Lambda^2$, at which the 
theory classicalizes, exist always, whereas for $\Box \gg \Lambda^2$ a Wilsonian mechanism might be at work. Accordingly, Fierz--Pauli theory on FRW has always a finite 
classicalitzation window \cite{Dvali:2011nj}.

The original classicalization proposal embraced interacting field theories
over Minkowski spacetime. Here, we showed explicitly that the classicalization
paradigm extends to free field theories on curved spacetimes.  

~\\

\textit{Summary.}---In this Letter, we have shown that generic relevant deformations of Einstein's gravity 
theory feature strong coupling phenomena (among the additional degrees 
of freedom) that originate from energy-momentum self-sourcing, corresponding
to a feedback through nonlinear terms. 
Moreover, this kind of self-protection follows precisely the recently proposed 
classicalization paradigm, however, extending its domain to include
free field dynamics on curved backgrounds. 

We have demonstrated explicitly that the classicalization window is
open for Fierz--Pauli like deformations of
gravity on FRW spacetimes. A Wilsonian mechanism could only close it partially (at high energies), then leading to a finite classicalization window.
There, classicalization proceeds 
through a strong coupling regime that triggers the massive
decay of the original vacuum and signals the formation of the largest
possible classicalon --- the new background spacetime. 

The consistency of relevant deformations on arbitrary background
is thus implied --- a necessary prerequisite for a nonlinear
completion. In this respect, classicalization might present 
a self-protection mechanism that stabilizes the theory at the 
nonlinear level, provided the classicalization scale always beats
the characteristic scale for unitarity violation. 

The authors would like to thank 
Gia Dvali,
Michael Kopp,
Parvin Moyassari,
and
Florian Niedermann
for helpful and inspiring discussions.
D.D.D.~acknowledges gratefully the hospitality of the Arnold Sommerfeld Center and the Excellence Cluster Universe.
The work of S.H.~was supported by the DFG cluster of excellence `Origin and Structure of the Universe'
and by TRR 33 `The Dark Universe.'

\end{document}